# Small oscillations of an electric dipole in the presence of a uniform magnetic field

L A del Pino

Departamento de Física, Universidad de Antofagasta, Campus Coloso, Antofagasta, Chile.

Luis.delpino@uantof.cl

**Abstract**. In this letter we study the classical motion of an electric dipole in the presence of a uniform magnetic field in the approximation of small oscillations. The normal modes of oscillations are obtained and propose a criterion of applicability of the method of small oscillations. In our opinion, the resolution of such examples is useful for training students in science and engineering.

## 1. Introduction

Several potentials are used to model the movement of particles in Nature. An interesting behavior is seen when the particles move in a minimum of the potential. This behavior is so-called small oscillations. This method is the most common approximation that we find in the literature [1,2].

Several authors have paid much attention in a great variety of interesting details of physical problems [3,4], in almost of them it is possible to find reference to small oscillations as a curiosity, but there is not included any discussion about the validity of the approximation.

Here, we make an effort to provide a criterion to apply correctly the approximation of small oscillations in an electric dipole in the presence of a magnetic field. This case is a two-body problem and its exact analysis requires a difficult treatment and further useful understanding, which is obtained by approximations. From this, the relevance of the current discussion is stemmed.

Classical treatment constitutes an interesting approach, although a deep discussion of the present topic requires a quantum formulation. The present discussion can be useful to introduce an approach to facilitate the understanding of concepts about the problem of dipoles in a magnetic field at the senior undergraduate level, the discussion about the validity of the small approximation is independent on the level, and it is possible to do it to students of science and engineering.

In this letter, the problem of an electric dipole in a magnetic field in the Troncoso and Curilef [3] formulation is chosen to derive the main equations of motion. In this line, a criterion is proposed to validate the approximation of small oscillations, and the most important question that is tried to solve is: "What is the range of the energy, where the approximation is valid?"

## 2. Normal modes of oscillations.

Troncoso and Curilef derived [3], the classical motion of a rigid electric dipole in the presence of a uniform magnetic field, $\vec{B} = B\,\vec{e_3}$, which has two integrals of motion:

$$E = \frac{\mu}{2}\dot{r}^2 + \frac{1}{2M}\left|\frac{e}{c}\vec{B}\times\vec{r} - \vec{C}\right|^2 - \frac{e^2}{\kappa|\vec{r}|} \qquad (1)$$

$$\vec{C} = M\dot{\vec{R}} + \frac{e}{c}\vec{B}\times\vec{r} \qquad (2)$$

Where $\mu = \frac{m}{2}$ is the reduced mass; M = 2m is the total mass; $\vec{r} = \vec{r_2} - \vec{r_1}$ is the relative position; $\dot{\vec{R}}$ is the velocity of the centre of mass; E is the energy; and $\vec{C}$ is a constant vector. Let's turn to dimensionless variables. Let $\omega_c = \frac{eB}{Mc}$ the cyclotron frequency; $\tau = \omega_c t$ as the dimensionless time; $\vec{\rho} = \frac{\vec{r}}{d}$; $\vec{\xi} = \frac{\vec{R}}{d}$; $\varepsilon = \frac{E}{E_0}$; with $E_0 = \frac{1}{2}Md^2\omega_c^2$ and $\varepsilon_c = \frac{e^2}{E_0\kappa d}$, where $d$ is the length of the dipole. In these new variables (1) and (2) becomes:

$$\varepsilon + \varepsilon_c = \frac{1}{4}\dot{\rho}^2 + \left|\vec{e_3}\times\vec{e_\rho} - \vec{C}\right|^2 \qquad (3)$$

$$\vec{C} = \dot{\vec{\xi}} + \vec{e_3}\times\vec{e_\rho}. \qquad (4)$$

Consider these two cases correspond to two possible modes of oscillation. Initially, the dipole lies at rest in the perpendicular plane to the magnetic field (mode I) and the dipole lies at rest in the direction parallel to the field (mode II). In the first case of (3) and (4) we have:

$$\varepsilon + \epsilon_c = \frac{1}{4}\dot{\varphi}^2 + 2(1 - \cos\varphi), \qquad (5)$$

where $\varphi$ is the polar angle and $\dot{\varphi} = \dot{\rho}$. Equation (5) is the well-known equation used for the simple pendulum whose exact period is $T_1 = 2K\left(\sin^2\left(\frac{\varphi_0}{2}\right)\right)$; where $K(x)$ is the complete elliptic integral of the first kind and $\varphi_0$ is the turning point. To analyze, the behaviour in the small oscillations is sufficient to do in (5), with $\cos\varphi \approx 1 - \frac{1}{2}\varphi^2$ and $\varphi = \sqrt{2}\,\eta_1$; thus (5) becomes:

$$\varepsilon + \varepsilon_c = \frac{1}{2}\dot{\eta}_1^2 + 2\,\eta_1^2. \qquad (6)$$

Equation (6) is the equation for harmonic oscillator of the dimensionless frequency, $\omega_1^{so} = 2$; and $\eta_1$ is the normal coordinate, whose normal period is $T_1^{so} = \pi = 2K(0)$.
In the second case (3) becomes:

$$\varepsilon + \varepsilon_c = \frac{1}{4}\dot{\theta}^2 + \sin^2\theta \qquad (7)$$

where $\theta$ is the azimuthal angle. This case corresponds to the bound-and-trapped states considered in the literature [3,5]. The exact period of the equation (7) is $T_2 = 2K(\sin^2\theta_0)$. If in equation (7) we do the approximation, $\sin\theta \approx \theta$ and $\theta = \sqrt{2}\,\eta_2$, we obtained:

$$\varepsilon + \varepsilon_c = \frac{1}{2}\dot{\eta}_2^2 + 2\,\eta_2^2 \qquad (8)$$

which is again the equation of a harmonic oscillator of frequency, $\omega_2^{so} = 2$; and $\eta_2$ is the normal coordinate whose normal period is $T_2^{so} = \pi$. Returning to the dimensional variables, the periods corresponding to each normal mode of oscillation are $T_1^{so} = T_2^{so} = \frac{\pi}{\omega_c}$.

Finally, consider the issue of the applicability of the method to small oscillations. That is, try to answer the question: *what is the energy range where the approximation represents the studied model?*

To answer, this question considers the following: In general the normal periods for each normal mode of oscillation are estimated, from parameters that define the Hamiltonian of the system. These parameters are known with some degree of uncertainty that spreads to the calculation of these periods. Let $\Delta T_s^{so}$ be the estimating error of the corresponding normal period, where $s$ is the number of degree of freedom and $\delta_s(E) = |T_s - T_s^{so}|$ the absolute deviation between the exact period and corresponding normal period. Then the inequality:

$$\delta_s \leq \Delta T_s^{so}. \qquad (9)$$

Which means, within the margin of error imposed by $\Delta T_s^{so}$, it is impossible to distinguish the exact period of the system and its corresponding normal period. Therefore, the range of energy values that satisfy (9) is the energy range where the approximated results coincide with the exact one of the system.

On this case, the relative error of the normal period estimation is given by the relative error of the cyclotron frequency estimation, that is, $\frac{\Delta T_s^{so}}{T_s^{so}} = \frac{\Delta \omega_c}{\omega_c}$. In a system of units where the total mass of the dipole and its length is taken equal to unity, the relative error of the cyclotron frequency estimation is determined by the relative error of the dipole moment $|\vec{P}|$ measurement, for a fixed magnetic field. This error, for polar linear molecules [6] (HF, HCL, Hbr) is typically 1%. The solutions of (9) for this relative error are:

$\varphi_0 \leq 0.4 \, rad$ and $\theta_0 \leq 0.2 \, rad$ respectively (see figure 1). The final answer to this problem of small oscillations is:

Mode I

$$\eta_1 = \frac{\sqrt{2}}{2}\varphi, T_1^{so} = \frac{\pi}{\omega_c}, \vec{\dot{R}}_0 = 0, \vec{B} \cdot \vec{P_0} = 0, -0.8\omega_c \leq \dot{\varphi}_0 \leq 0.8\omega_c$$

Mode II

$$\eta_2 = \frac{\sqrt{2}}{2}\theta, T_2^{so} = \frac{\pi}{\omega_c}, \vec{\dot{R}}_0 = 0, \vec{B} \times \vec{P_0} = 0, -0.4\omega_c \leq \dot{\theta}_0 \leq 0.4\omega_c$$

Where $\vec{\dot{R}_0}$ is the initial velocity of the center of mass; $\vec{P_0}$ is the initial dipole moment; and $\dot{\varphi}_0, \dot{\theta}_0$ are the initial angular velocities of the dipole.

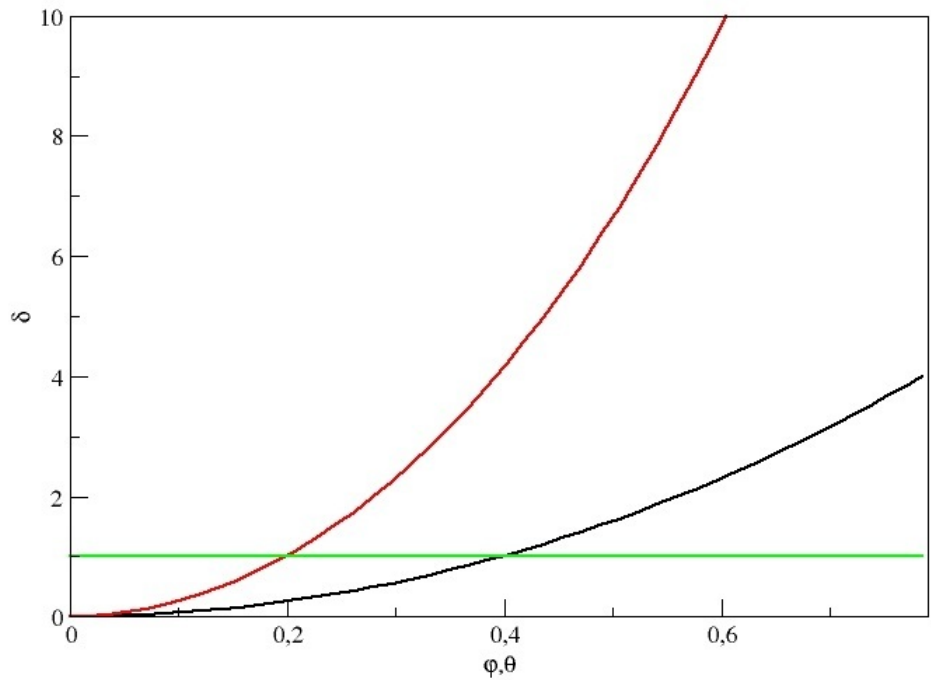

Figure 1. Numerical solution of (9). $\delta$ is expressed in percent. The black line corresponds to model I and the red line to the mode II.